# Preferred corrosion pathways for oxygen in $Al_2Ca$ – twin boundaries and dislocations

Nicolas J. Peter,[1,2] Daniela Zander,[3] Xumeng Cao,[4] Chunhua Tian,[1] Siyuan Zhang,[1] Kui Du,[4] Christina Scheu[1] and Gerhard Dehm[1,a]

[1]*Max-Planck Institute for Iron Research, Max-Planck-Strasse 1, 40237 Düsseldorf, Germany*

[2]*Institute of Energy and Climate Research (IEK-2), Forschungszentrum Jülich GmbH, 52425 Jülich, Germany*

[3]*Chair of Corrosion and Corrosion Protection, Division of Materials Science and Engineering, RWTH Aachen University, Intzestrasse 5, 52072 Aachen, Germany*

[4]*Shenyang National Laboratory for Materials Science, Chinese Academy of Science, 110016 Shenyang, China*

With an ongoing discussion on the oxygen diffusion along crystal defects remaining, it is difficult to study this phenomenon in Al containing intermetallic materials due to its rapid and passivating oxide formation. We report here the observation of enhanced oxygen diffusion along crystal defects, i.e. dislocations and twin boundaries, in the C15 $Al_2Ca$ Laves phase and how the presence of oxygen induces structural changes at these defects. Three main phases were identified and characterized structurally by aberration-corrected, atomic resolution scanning transmission electron microscopy, analytically by energy dispersive X-ray spectroscopy and electron energy loss spectroscopy. Unlike the C15 bulk phase, the twin boundary and dislocation transformed into a few nanometer wide amorphous phase, which depletes in Al and Ca but is highly enriched in oxygen. The dislocation even shows coexistence of the amorphous phase with a simple Al-rich A1 fcc phase. This A1 phase only depletes in Ca, not in Al (Al remains at bulk concentration), and is also enriched in oxygen. The Al-rich A1 phase is coherent with the C15 matrix. Electron energy loss spectroscopy revealed the amorphous phase to be $Al_2O_3$. We thereby show as one of the first studies that oxygen diffusion along crystal defects, especially also at the twin boundary can induce the formation of an amorphous oxide along themselves. The identification of oxygen-induced transformation at strained defects has to be considered when the material is exposed to air during plastic deformation at elevated temperatures.

---

a) Author to whom correspondence should be addressed. Electronic mail: dehm@mpie.de

**MAIN TEXT**

Magnesium alloys belong to the lightest structural metals but exhibit low room temperature ductility and creep resistance. Al and Ca additions to such alloys were found to be beneficial in order to improve their creep resistance [1], strength and particularly their ductility [2], as well as their corrosion properties [3]. In this context, the formation of the intermetallic C15 $Al_2Ca$ Laves phase and its positive influence on ductility is recently discussed e.g. for dilute Mg-Al-Ca alloys [3]. However, there are only limited studies on the properties of the isolated $Al_2Ca$ phase in terms of its mechanical behavior [4]. Besides, there is still an ongoing discussion whether or not crystal defects, and in particular dislocations and twin boundaries, provide a fast diffusion pathway for oxygen species in alloys and intermetallic materials [5–9]. These defects could thereby critically influence oxidation rates of such materials, while being necessary for carrying plastic deformation. Specific attention is paid to the understanding of the transport mechanisms influenced by space charge regions, (interfacial) strain, segregation and electronic transfer along interfaces.

So far, very limited studies on the diffusion of oxygen in the intermetallic C15 $Al_2Ca$ phase is available. Even for fcc aluminum and its alloys investigations on dislocation and twin boundary driven oxygen diffusion mechanisms are scarce due to the rapid formation of the surface oxide. Only recently, the diffusion of interstitial oxygen in fcc aluminum has been studied using first-principles [10]. In contrast to most fcc metals, oxygen sits on the tetrahedral interstitial and not on the octahedral sites. However, the calculated oxygen diffusion coefficient is on the same order of magnitude of the diffusion coefficient in other fcc metals and has been calculated to be about $1.6 \times 10^{-23}$ m²/s at 273 K and $4.5 \times 10^{-11}$ m²/s at 873 K [10]. Notably, the introduction of such an oxygen interstitial changed the cell volume (strain) by about 1.5 %.

Alumina scales forming on crystalline interfaces and surfaces were already studied extensively in dry and humid oxygen atmospheres since the 19th century [11–15]. It was reported that the formation of either amorphous or crystalline oxide can occur based on thermodynamic (energetic) arguments [16]. For thin films, the sum of the surface energy of an amorphous oxide and interface energy between metal and amorphous metal oxide is lower than the oxide's bulk energy. At the point where the summed up interfacial contributions equal the oxide's bulk energy, a critical thickness is reached to form crystalline metal oxide. This critical thickness of metal oxides forming on metal surfaces has been calculated to be one or two atomic layers for (111) Al substrates up to about 4 nm for (110) Al at room temperature, which can increase to about 0.5 nm and 7 nm at 900 K [16]. In addition, it was assumed that the low diffusion coefficient of aluminum species at low temperatures ≤ 573 K result in an amorphous oxide layer with a limited but uniform thickness whereas at higher temperatures > 673 K the initial amorphous oxide layer transforms into $\gamma$-$Al_2O_3$ after prolonged oxidation [14]. Another experimental study confirmed this transformation on an epitaxially grown thin film [17]. Only recently, ab initio simulations on the formation and stability of amorphous, $\gamma$-$Al_2O_3$ and $\alpha$-$Al_2O_3$ were published and related to the thermodynamic driving forces behind the initial phase selection among amorphous and crystalline structures growing on a fcc (111) aluminum substrate. It was found that $\gamma$-$Al_2O_3$ becomes more stable than the amorphous polymorph for layer thicknesses down to ~1 nm at 600 K, whereas below 1 nm the amorphous layer becomes more stable[18]. However, there is no information on the oxygen transport as well as the nucleation and growth mechanism of the metal oxides formed at dislocations and twin boundaries and in particular of the $Al_2Ca$ Laves phase. As a consequence, the aim of the present study is to discuss the influence of dislocations, and twin boundaries acting as 1) fast oxygen pathways and 2) preferred nucleation sites for the oxygen-induced formation of amorphous $Al_2O_3$ for the $Al_2Ca$ Laves phase.

The used material was cast at a target Al:Ca ratio of 2:1 to obtain the $Al_2Ca$ alloy and was subsequently heat treated at 600 °C for 24 hours under Argon atmosphere. Cast alloys were not quenched but cooled in the furnace under argon atmosphere. However, processing under these conditions and considering a heat of formation of -1657 kJ/mol for ($\gamma$-)$Al_2O_3$[19] and of -635 kJ/mol for CaO[20] at room temperature and a pressure of one atmosphere still makes oxygen interaction with the $Al_2Ca$ surface as well as defects close to the surface even for low remaining oxygen partial pressures very likely. Metallographically prepared samples from the cast alloy were investigated. Macroscopic/bulk chemistry and structure were investigated by inductively coupled plasma optical emission spectroscopy (ICP-OES) and scanning electron microscopy (SEM) – back scattered electron (BSE) imaging, respectively. High magnification atomic structure characterization was performed using aberration-corrected (scanning) transmission electron microscopes (S/TEM) at 300 kV. Energy-dispersive X-ray spectroscopy (EDS) and electron energy loss spectroscopy (EELS) were used inside the STEM to study local chemistry and for phase identification, respectively. Strain determination is based on experimentally acquired STEM-HAADF micrographs, the exact method being described in detail by K. Du et al.[21]. STEM multislice image simulations are based on the PRISM algorithm[22]. Detailed experimental information of all characterization techniques is reported in the supplementary information.

BSE imaging of the cast and annealed material revealed the presence of a two-phase microstructure (Figure 1a), i.e. a dominant matrix phase ($Al_2Ca$) and a fine skeletal phase ($Al_4Ca$) along the material's grain boundaries. Within individual grains of the matrix phase straight features (annealing twins) are observed, which are highlighted in Figure 1a by white arrows. Closer observation under electron channeling contrast (ECCI) two-beam conditions (Figure 1b) revealed the presence of two twins (dashed lines), which follow the <011> crystal orientation and appear decorated by line defects fading into the bulk. This fading contrast is likely originating from disconnections at the twin boundary and their associated line defects according to some further TEM investigations which are not discussed in this study. Additionally, a second phase $Al_4Ca$ particle has been captured (black contrast in Figure 1b). The location of the S/TEM lamella lift-out, perpendicular to the <011> orientation is indicated in Figure 1b as well by a solid white line. The lower twin boundary 2 in Figure 1b as well as the second phase particle can be seen in one part of the extracted S/TEM lamella (Figure 1c), while two dislocations close to the surface are visible in another part of the same lamella (Figure 1d). With a close look, these HAADF micrographs reveal that the lower twin boundary of the twin has some dark contrast ranging about 500 nm from the surface into the bulk along the twin boundary (Figure 1c). This contrast is not seen at the upper twin boundary, but is also seen at dislocation 2 close to the surface in Figure 1d. The overall chemical bulk composition of the heat treated alloy was determined by ICP-OES to be 64.3 at. % aluminum and 35.7 at. % calcium, which is close to the targeted composition. Local chemical analysis at higher magnification by STEM-EDS revealed good agreement of the identified phases to the expected $Al_2Ca$ matrix and the $Al_4Ca$ skeletal phase nominal compositions. The results are summarized in Figure 1e. Finally, the $Al_2Ca$ matrix phase revealed an atomic arrangement by HAADF-STEM imaging in [011] zone axis orientation, which fits to the expected C15 Laves phase crystal structure (Figure 1f). Brighter dots correspond to Ca atomic columns, while less bright dots are Al atomic columns. A multislice STEM image simulation of the C15 structure is presented as inset for reference.

Detailed analysis of the twin inside the bulk reveals two perfectly intact twin boundaries (Figure 2a). The TEM image's fast Fourier transform is depicted in the inset, which shows the characteristic mirror symmetry and {111} as well as {200} twin reflections. Atomic resolution HAADF-STEM of the lower twin boundary in Figure 2a has been performed at different locations, i.e. in the bulk with a perfect twin boundary arrangement (Figure 2b) and close to the surface with a distinct disruption of the atomic arrangement by an amorphous area at the former location of the twin boundary (Figure 2d), [{111} lattice planes are indicated for reference]. The latter location was accompanied by EDS analysis, which revealed the amorphous region to contain less Al and Ca, but up to 30 at. % of oxygen (Figure 2e). This intergranular region is widest at the surface and narrows down into the bulk until it fades into the perfect twin boundary atomic arrangement described above. We estimated a simple diffusion coefficient (D) for oxygen along the twin boundary to be $D \approx 1.4 \times 10^{-18}$ m²/s knowing the diffusion length (x = 500 nm) and annealing time (t = 24 h) according to $D = x^2/t$. For the former location, and thus the perfect twin boundary arrangement, strain analysis was performed (Figure 2c) revealing that the lattice planes next to the twin boundary plane gradually reach a compressive strain of up to 1.09 ± 0.18 % (a gaussian distribution is plotted as purple dotted line to guide the eye). Compressive strains were characterized in literature e.g. for γ-Fe channels between κ-carbide precipitates to be on the order of 5 %[23]. The detected maximum compressive strain of 1 % of the twin boundary appears reasonable, as the just slight distortion of the atomic arrangement at the twin boundary is expected to lead to elastic strains only. It is speculated that the detected strain at the interface causes the preferred pathway for enhanced oxygen transport along the twin boundary as we were able to exclude segregation effects by high magnification, analytical STEM. The presence of oxygen at the twin boundary, potentially in combination with the mechanical (stress) strain, allows for the heterogeneous nucleation of the amorphous interface phase from the boundary by the chemomechanical stimulus.

Similar behavior is found for dislocation 2 in Figure 3a. The dark contrast along the dislocation line at low magnification is again found to mainly be an amorphous region (dislocation 2 in Figure 3a). However, inside this region, crystalline areas were identified consisting of a different crystal structure as compared to the C15 $Al_2Ca$ matrix phase. In fact, a simple fcc (A1) crystal structure was formed coherently with the C15 adjacent matrix phase lattice and in coexistence to the amorphous areas (Figure 3b). There are instances where this newly formed A1 phase even contains a twin in itself. In another dislocation (dislocation 1 in Figure 3a) located further into the bulk and not connected to the surface, there is no such dark contrast as is the case at dislocation 2, but a disordered structure of only few atomic unit cells (Figure 3d, disruption of lattice planes indicated by arrow). The continuation of the atomic arrangement across the feature is distorted, i.e. lattice planes are displaced by half a lattice plane distance, as indicated by white dashed lines and an arrow. Local chemical analysis across the amorphous and A1 crystalline areas of dislocation 2 in Figure 3b was performed by STEM-EDS spectral imaging (mapping) and quantitative analysis in terms of molar fraction of the individual elements Al, Ca and O (Figure 3c). It was found that in the crystalline region the molar fractions of Al and Ca do not change much from the bulk ($Al_2Ca$), however, oxygen is slightly enriched. By contrast, in the amorphous regions the molar fractions of Ca and especially Al are much reduced, while strong oxygen enrichment is found. From the molar fraction this enrichment roughly fits to a $M_2O_3$ (M being metal), indicative for $Al_2O_3$. The extracted EDS concentration profiles are provided in the supplementary material. EELS measurements from the amorphous region fit well to the O-K edge of amorphous $Al_2O_3$ [17] (Figure 3e), however, the energy resolution does not allow to resolve the subtle differences between amorphous- and gamma $Al_2O_3$. In any case, the differences in chemical composition of the identified states allows for a transformation sequence to be postulated. From the initial C15 $Al_2Ca$ phase Ca is the first element to deplete and form the A1 phase, while in a subsequent step Al is additionally lost leading to the formation of the amorphous $Al_2O_3$ phase during further oxygen incorporation.

It appears that strain containing defects in $Al_2Ca$ including dislocations and twin boundaries are the reason for fast oxygen diffusion, and as a consequence, for oxygen-induced structural changes. This is supported by comparison of the appearance of the amorphous regions along dislocation 2 and the twin boundary. While comparable in length, the dislocation even being slightly longer, the twin boundary has a wide amorphous region at the surface, which narrows down along its length, while the dislocation appears to have a constant width of the amorphous region along its entire length. This suggests that the higher strain at the dislocation core (typically up to about 10 % for edge dislocations) leads to even higher diffusion rates for oxygen than the twin boundary.

In conclusion, we show the oxygen diffusion-induced formation of amorphous $Al_2O_3$ along crystal defects, i.e. a dislocation and a twin boundary. At the dislocation a coexistence of an oxygen containing Al rich A1 phase and the amorphous alumina phase has been observed. From the chemical composition of the two phases, we believe that a loss of Ca precedes the reduction of Al within the structurally transformed region, suggesting a transformation sequence of C15 $Al_2Ca + O_2 \rightarrow$ A1 Al(O) phase $\rightarrow$ amorphous $Al_2O_3$. Understanding the role of oxygen diffusion pathways is crucial for using the material reliably, as already small amounts of oxygen can lead to structural transformations along defects.

## ACKNOWLEDGMENTS

The authors acknowledge financial support by the Deutsche Forschungsgemeinschaft (DFG) through the projects B01, B05, and B06 of the SFB1394 Structural and Chemical Atomic Complexity – From Defect Phase Diagrams to Material Properties, project ID 409476157. The authors cordially thank Hauke Springer and Leandro Tanure (RWTH Aachen) for preparation of the annealed $Al_2Ca$ alloy. In addition, the authors like to acknowledge Uzair Rehman and Supriya Nandi (Max-Planck Institut für Eisenforschung GmbH) for assistance with the $Al_2Ca$ sample and ECC imaging, respectively.

**FIGURES**

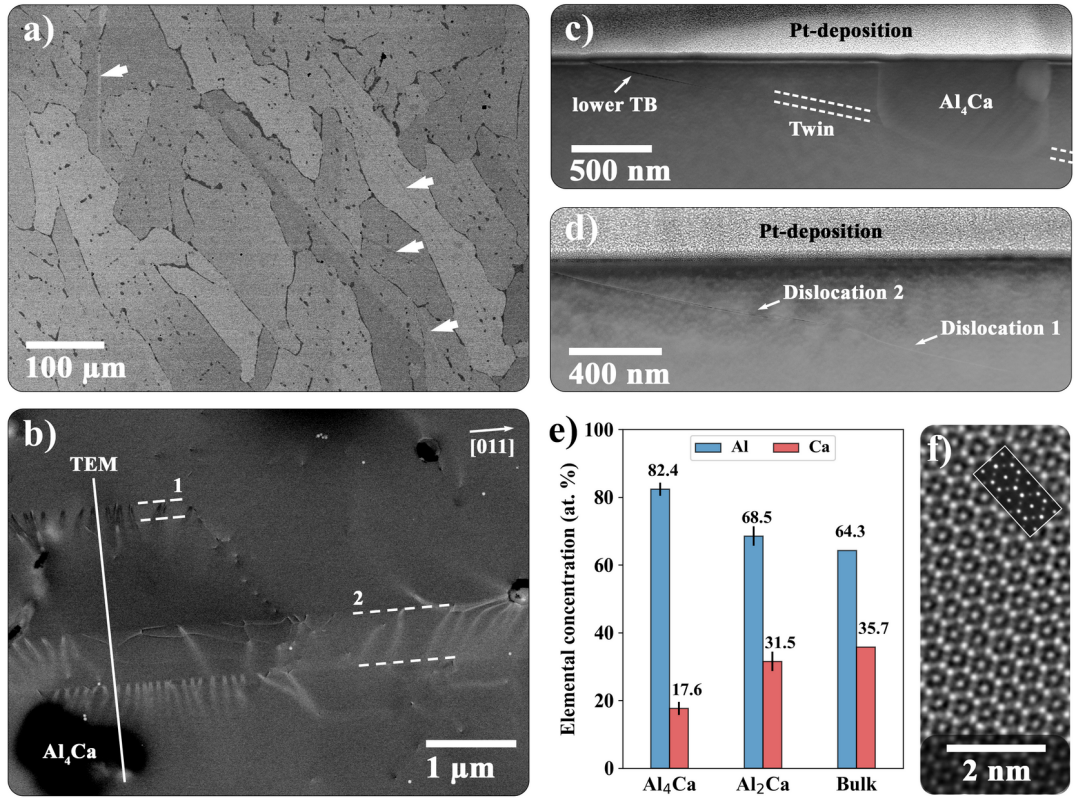

Figure 1: Overview of the as-cast and heat treated $Al_2Ca$ alloy. BSE overview image of the annealed microstructure with twins being indicated by white arrows (a). ECC image of the region of interest showing several microstructural features including two twins 1 & 2 and a second phase $Al_4Ca$ (b). The [011] crystal direction and the location of a S/TEM lift-out are indicated. STEM-HAADF micrographs of a twin and a two dislocations are presented in Figures (c) and (d), respectively. Note the dark contrast of the lower twin boundary in (c) and at dislocation 2 in (d). The chemical composition of the bulk and the two present phases, as determined by ICP-OES and STEM-EDS, are displayed in (e), while the atomic arrangement is provided in the atomically-resolved STEM-HAADF micrograph in (f). For reference, a multislice STEM image simulation result of the $Al_2Ca$ C15 Laves phase in [011] zone axis orientation is shown as inset.

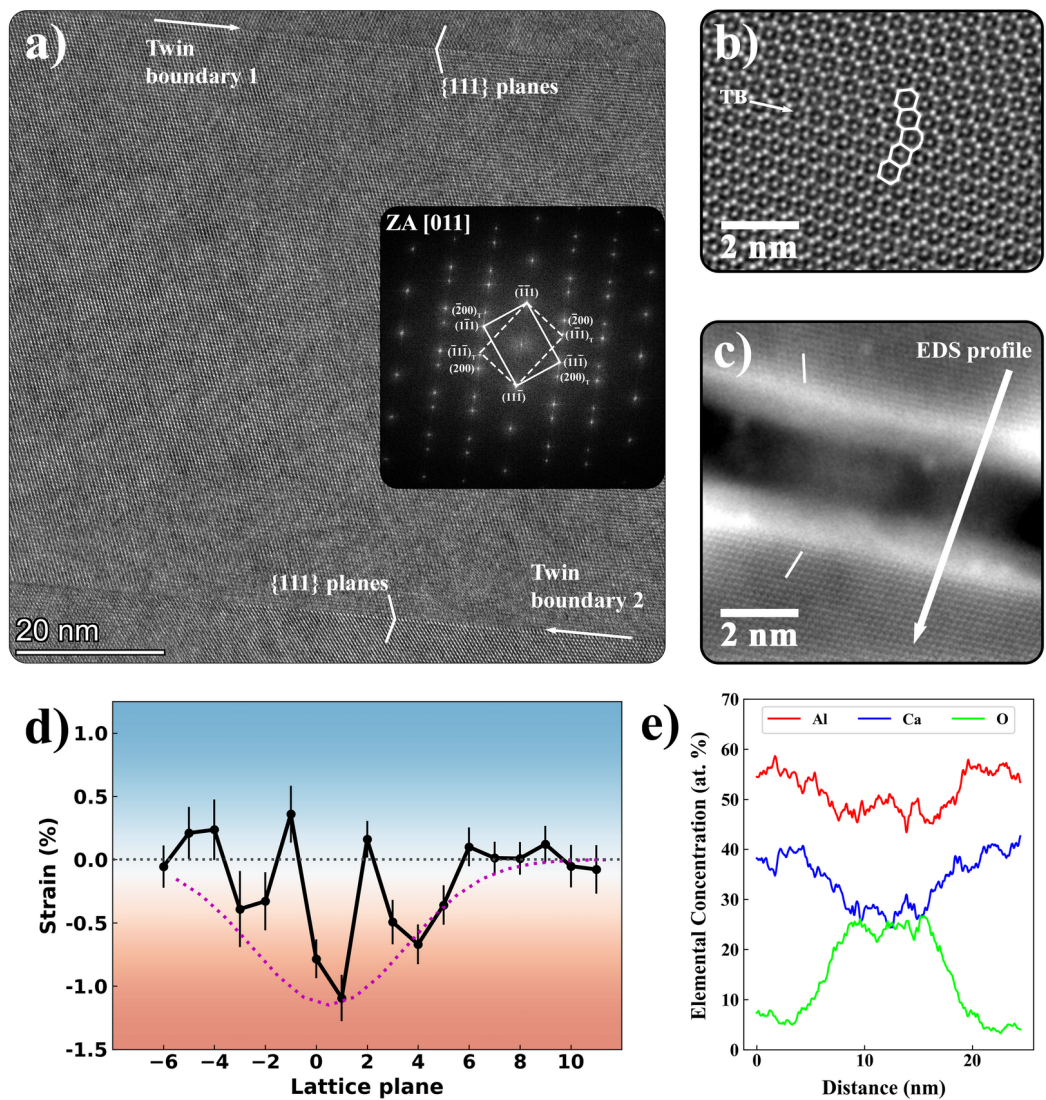

Figure 2: Characterization of oxygen enrichment induced changes at the twin boundary. In the bulk, the twin is composed of two perfectly arranged twin boundaries as seen in the high-resolution TEM image (a). The inset reveals the characteristic mirror symmetry and {200} as well as {111}-twin reflections in [011] zone axis orientation. The perfect atomic arrangement of the twin boundary is captured in the atomically-resolved HAADF -STEM micrograph (b), while the amorphized twin boundary region does not contain any atomic order anymore (c). Note that the amorphization only appears at twin boundary 2 in (a), but not in twin boundary 1. Strain of lattice planes (d) across the twin boundary is presented for the intact twin boundary in (b) and reveals about 1 % compressive strain in the close vicinity of the twin boundary. The EDS profile (e) across the amorphized twin boundary portion (d) shows the enrichment of the structurally transformed regions with oxygen and the depletion of Al and Ca.

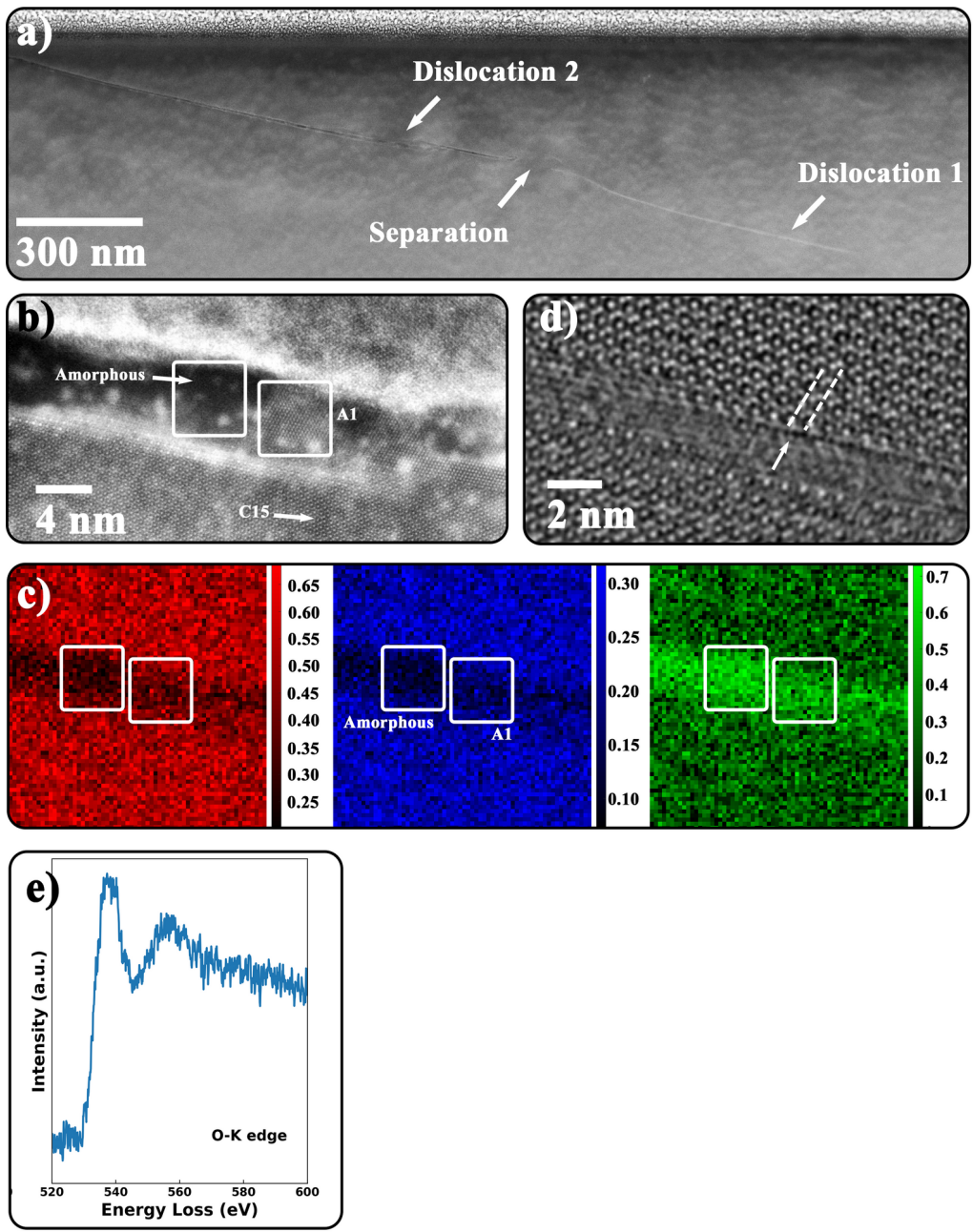

Figure 3: Characterization of oxygen enrichment induced changes at the (sub)surface dislocations. Two dislocations were identified in the HAADF-STEM micrograph in (a), i.e. dislocation 2 running into the surface, while dislocation 1 follows after a short separation distance. Dislocation 2 reveals a distinct structural arrangement in its oxygen enriched region by HAADF-STEM (b). There are amorphous regions coexisting with simple fcc (A1) crystalline regions, both highlighted by white squares. Chemical analysis of this exact area by quantitative STEM-EDS spectral imaging analysis (c) shows that amorphous region reveal a close to $Al_2O_3$ molar composition (color scale in molar fraction), while the crystalline part is close to the $Al_2Ca$ composition. Dislocation 1, in contrast to dislocation 2, is only disordered for few atomic unit cells, as seen in the HAADF-STEM micrograph in (d). The continuation of the lattice planes is indicated with a white dashed line. (e) EELS fine structure of the O-K edge.